\documentstyle[12pt,fleqn]{article}
\input epsf

\def\singlespace {\smallskipamount=3.75pt plus1pt minus1pt
                  \medskipamount=7.5pt plus2pt minus2pt
                  \bigskipamount=15pt plus4pt minus4pt
                  \normalbaselineskip=15pt plus0pt minus0pt
                  \normallineskip=1pt
                  \normallineskiplimit=0pt
                  \jot=3.75pt
                  {\def\smallskip {\vskip\smallskipamount}}
                  {\def\medskip   {\vskip\medskipamount}}
                  {\def\bigskip   {\vskip\bigskipamount}}
                  {\setbox\strutbox=\hbox{\vrule
                    height10.5pt depth4.5pt width 0pt}}
                  \parskip 7.5pt
                  \normalbaselines}
\def\middlespace {\smallskipamount=5.625pt plus1.5pt minus1.5pt
                  \medskipamount=11.25pt plus3pt minus3pt
                  \bigskipamount=22.5pt plus6pt minus6pt
                  \normalbaselineskip=22.5pt plus0pt minus0pt
                  \normallineskip=1pt
                  \normallineskiplimit=0pt
                  \jot=5.625pt
                  {\def\smallskip {\vskip\smallskipamount}}
                  {\def\medskip   {\vskip\medskipamount}}
                  {\def\bigskip   {\vskip\bigskipamount}}
                  {\setbox\strutbox=\hbox{\vrule
                    height15.75pt depth6.75pt width 0pt}}
                  \parskip 11.25pt
                  \normalbaselines}
\def\doublespace {\smallskipamount=7.5pt plus2pt minus2pt
                  \medskipamount=15pt plus4pt minus4pt
                  \bigskipamount=30pt plus8pt minus8pt
                  \normalbaselineskip=30pt plus0pt minus0pt
                  \normallineskip=2pt
                  \normallineskiplimit=0pt
                  \jot=7.5pt
                  {\def\smallskip {\vskip\smallskipamount}}
                  {\def\medskip   {\vskip\medskipamount}}
                  {\def\bigskip   {\vskip\bigskipamount}}
                  {\setbox\strutbox=\hbox{\vrule
                    height21.0pt depth9.0pt width 0pt}}
                  \parskip 15.0pt
                  \normalbaselines}

\include{dspace12}

\def\be{\begin{equation}}
\def\ee{\end{equation}}

\def\sect #1{\setcounter{equation}{0}}

\begin{document}
\doublespace

\begin{center}
\Large {\bf Gravitational Collapse}
\end{center}
\vspace{1.0in}
\vspace{12pt}
\begin{center}
{\large{Pankaj S. Joshi\\
Tata Institute of Fundamental Research\\
Homi Bhabha Road, Bombay 400005, India.\\
}}
\end{center}
\vskip 1 in
\centerline{\bf ABSTRACT}
\medskip

\noindent
We review here some recent developments on the issue of final
fate of gravitational collapse within the framework of Einstein
theory of gravity. The structure of collapsed object is discussed
in terms of either a black hole or a singularity having causal connection
with outside universe. Implications for cosmic censorship are
discussed.

\bigskip

\noindent {\bf 1. Introduction}
\smallskip

When a massive star, more than a few solar masses, has exhausted its
internal nuclear fuel, it is believed to enter
the stage of an endless gravitational collapse without having any final
equilibrium state. According to the Einstein theory of gravitation, the
star goes on shrinking in its radius, reaching higher and higher densities.
What would be the final fate of such an object according to the general
theory of relativity? This is one of the central questions in
relativistic astrophysics and gravitation theory today. It has been
suggested that the ultra-dense object that forms as a result of collapse
could be a black hole in the space and time from which not even light rays
can escape. Alternatively, if the event horizon of gravity fails to cover the
final crunch, it could be a visible singularity which can causally interact
with the outside universe and from which emissions of light and
matter may be possible.

An investigation of this nature is of importance from both the
theoretical as well as observational point of view.
At the theoretical level, working out the final fate of collapse in
general relativity is crucial to the problem of cosmic censorship and
asymptotic predictability [1]; namely, whether the singularities forming
at the end point of collapse will be necessarily covered by the event
horizons of gravity. Such a cosmic censorship hypothesis remains fundamental
to the theoretical foundations of entire black hole physics, and
its numerous astrophysical applications which have been invoked
in past decades (e.g. the area theorem for black holes, laws of black hole
thermodynamics, Hawking radiation effect, predictability; and on
observational side, accretion
of matter by black holes, massive black holes at the center of galaxies etc).
On the other hand, existence of visible or naked singularities
would offer a new approach on these issues requiring modification
and reformulation of our usual theoretical conception on censorship.

Our purpose here is to discuss some of the recent developments
in this direction, examining the possible final fate of gravitational
collapse in general relativity. To investigate this issue, dynamical
collapse scenarios have been examined in the past decade or so for
many cases such as clouds composed of dust, radiation, perfect fluids,
or also of matter compositions with more general equations of
state (for references and details, see e.g. [2]).
We try to discuss some of these developments and the
implications of this analysis towards a possible formulation of cosmic
censorship are pointed out. Finally, the
open problems in the field are discussed and some concluding remarks try to
summarize the overall situation.

In Section 2, some features of spherically symmetric
collapse and basic philosophy on cosmic censorship are discussed.
It is the spherical collapse of a homogeneous dust cloud,
as described by the Oppenheimer$-$Snyder model [3], which led to the general
concept of trapped surfaces and black hole. The expectation is,
even when collapse is inhomogeneous or non-spherical, black holes must
always form covering the singularity implied by the trapped surface.
Some available versions of censorship hypothesis are discussed,
pointing out formidable difficulties for any possible proof.
It is concluded that the first major task is to formulate
a provable version of censorship conjecture; and that to achieve this,
a detailed and careful analysis of available gravitational
collapse scenarios is essential where the possible occurrence and physical
nature of the naked singularity forming is to be analyzed.
It is only such an examination of collapse situations which could tell us
what  features are to be avoided, and which ones to look for,
while formulating and proving any reasonable version of censorship.
This would also lead us to a better and more effective
understanding of the nature and occurrence of naked singularities in
gravitational collapse.

Towards this purpose, Section 3 reviews the phenomena of black hole and
naked singularity formation for gravitational collapse of
several different forms of matter. The discussion begins with the
radiation collapse models which provide
an explicit and clear idea on the nature and structure of the singularity
forming. Further generalizations of these results include forms of matter
such as perfect fluids, dust, and also general matter fields
satisfying the weak energy condition. The emerging pattern shows that
for a rather general form of equation of state and collapsing matter,
it is possible for a non-zero measure set of non-spacelike trajectories to
escape from the naked singularity which could form from a regular initial
data, and which could be powerfully strong for the growth of curvatures.
Non-spherical collapse and scalar field collapse are
also discussed here. Section 4 gives some concluding remarks
and future directions.

\bigskip

\noindent {\bf 2. Spherically Symmetric Collapse and Cosmic Censorship}
\smallskip

To understand the possible final fate of a massive gravitationally
collapsing object, we first outline here
the spherically symmetric collapse situation.
Though this represents an idealization, the advantage is one can
solve this case analytically to get exact
results when matter is taken in the form of a homogeneous dust cloud.
In fact, the basic motivation for the idea and theory of black holes
comes from this collapse model.

Consider a gravitationally collapsing spherical massive star.
We need to consider the interior solution for the object which will
depend on the properties of matter, equation of state, and the physical
processes taking place within the steller interior. However, assuming
the matter to be
pressureless dust allows to solve the problem analytically, providing
many important insights. Here the energy-momentum tensor
is given by $T^{ij}= \rho u^iu^j$, and one needs to solve the Einstein
equations for the spherically symmetric metric.
This determines the metric potentials,
and the interior geometry of the collapsing dust ball is given by,

\be
ds^2= -dt^2+ R^2(t)\left[ {dr^2\over 1-r^2}+ r^2 d\Omega^2 \right]
\ee
where $d\Omega^2=d\theta^2+ sin^2\theta d\phi^2$ is the metric on two-sphere.
The geometry outside is vacuum Schwarzschild space-time. The interior
geometry of the dust cloud matches at the
boundary $r=r_b$ with the exterior Schwarzschild space-time.

\begin{center}
\leavevmode\epsfysize=3.5 in\epsfbox{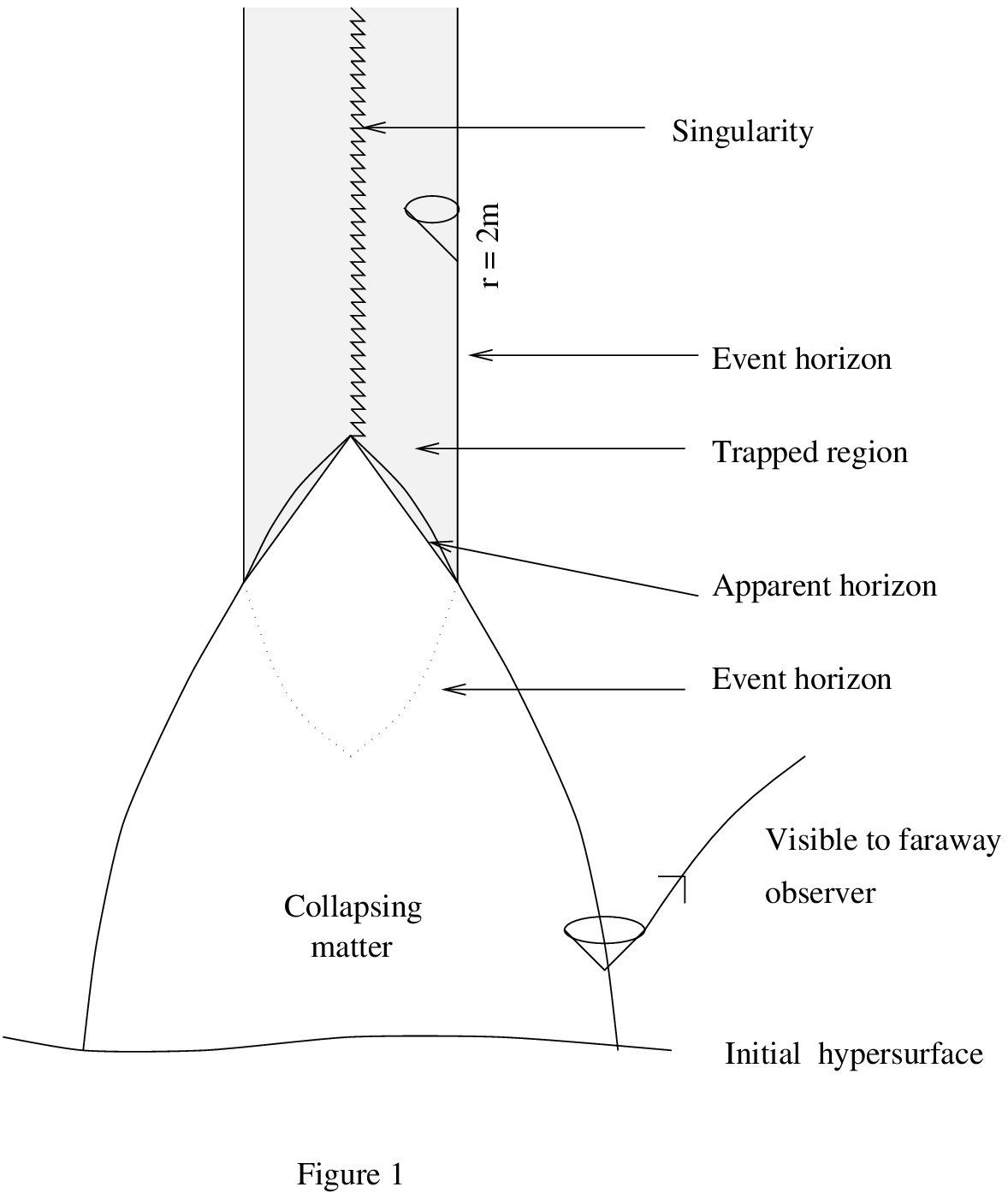}
\end{center}
\noindent {\small Fig. 1: The standard picture of spherically
symmetric homogeneous dust collapse.}

\bigskip

The basic features of such a collapsing, spherical, homogeneous dust
cloud configuration are given in Fig. 1.  The collapse is
initiated when the star surface is outside its Schwarzschild radius $r = 2m$,
and a light ray emitted from the surface of the star can escape to infinity.
However, once the star has collapsed below $r = 2m$, a black hole region
of no escape develops in the space-time, bounded by the event horizon
at $r =2m$.  Any point in this empty region
represents a trapped surface (which is a two-sphere in space-time)
in that both the outgoing and ingoing families of null geodesics emitted
from this point converge and hence no light comes out of this region.
Then, the collapse to an infinite density and
curvature singularity at $r =0$ becomes inevitable
in a finite proper time as measured
by an observer on the surface of the star. The black hole region in the
resulting vacuum Schwarzschild geometry is given by $0 < r < 2m$,
the event horizon being the outer boundary.
On the event horizon, the radial outwards
photons stay where they are, but all the rest are dragged inwards. No
information from this black hole can propagate outside $r=2m$ to
observers far away.
We thus see that the collapse gives rise to a
black hole in the space-time which covers the
resulting space-time singularity. The ultimate fate of the star
undergoing such a collapse is then an infinite curvature singularity
at $r = 0$, which is completely hidden within the black hole. No emissions or
light rays  from the singularity could go out to observer at infinity and the
singularity is causally disconnected from the outside space-time.

The question now is whether one could generalize these conclusions
on the occurrence of a space-time singularity in collapse
for non-spherical situation, and whether these are valid at least for small
perturbations from exact spherical symmetry.  It is known [4], using
the stability of Cauchy development in general relativity,
that the formation of trapped surfaces is indeed a
stable property when departures from spherical symmetry are taken into
account.  The argument essentially is the following: Considering a spherical
collapse evolution from given initial data on a partial Cauchy
surface $S$, we find the formation of trapped surfaces $T$
in the form of all the
spheres with $r < 2m$ in the exterior Schwarzschild geometry.  The stability
of Cauchy development then implies that for all initial data sufficiently
near to the original data in the compact region $J^{+} (S) \cap J^{-} (T)$,
the trapped surfaces still must occur.  Then, the curvature singularity of
spherical collapse also turns out to be a stable feature,
as implied by the singularity theorems, which show that
the closed trapped surfaces always imply the existence of a space-time
singularity under reasonable general conditions.

There is no proof available, however, that such a singularity will
continue to be hidden within a black hole and remain causally disconnected
from outside observers, even when the collapse is not exactly spherical
or when the matter does not have the form of exact homogeneous dust.
Thus, in order to generalize the notion of black holes to gravitational
collapse situations other than exact spherically symmetric homogeneous
dust case, it becomes necessary to rule out such naked or visible
singularities by means of an explicit assumption.  This is stated as the
{\it cosmic censorship hypothesis}, which essentially states that if
$S$ is a partial Cauchy surface from which collapse commences, then there
are no naked singularities to the future of $S$, that is, which could
be seen from the future null infinity. This is true for the spherical
homogeneous dust collapse, where the resulting space-time is future
asymptotically predictable and the censorship holds. Thus, the breakdown
of physical theory at the space-time singularity does not disturb prediction
in future for the outside asymptotically flat region.

What will be the corresponding scenario for other collapse situations,
when inhomogeneities, non-sphericity etc are allowed for? The answer to
this question is not known either in the form of a proof for the future
asymptotic predictability for general space-times, or of any
other suitable version of cosmic censorship hypothesis.
It is clear that the assumption of censorship in a suitable form is crucial
to basic results in black hole physics.  In fact, when one considers the
gravitational collapse in a generic situation, the very existence of
black holes requires this hypothesis [1].

If one is to establish the censorship by means of a rigorous proof,
that of course requires a much more precise formulation of the hypothesis.
The statement that result of a complete gravitational collapse must
always be a black hole and not a naked singularity, or all singularities of
collapse must be hidden in black holes, causally disconnected from observers
at infinity, is not rigorous enough. This is because, under completely general
circumstances, the censorship or asymptotic predictability is false as one
could always choose a space-time manifold with a naked singularity which would
be a solution to Einstein's equations if we define
$T_{ij} \equiv (1/8\pi)G_{ij}$.
Thus, at the minimum, certain conditions on the stress-energy tensor
are required, e.g. an energy condition.  However,
it turns out that to obtain an exact characterization of the  restrictions
one should require on  matter fields in order to prove a suitable version
of censorship hypothesis is an extremely difficult task and no
such specific conditions are available as yet.

The requirements in the black hole physics and general
predictability requirements
in gravitation theory have led to several different formulations of cosmic
censorship hypothesis.  The version known as the
{\it weak cosmic censorship} refers to
the asymptotically flat space-times and has reference to the null infinity.
Weak censorship, or asymptotic predictability, effectively postulates that
the singularities of gravitational collapse cannot influence events near
the future null infinity $\cal I^+$. If $S$ is the partial Cauchy surface on
which the regular initial data for collapse is defined, this is the requirement
that $\cal I^+$ is contained in the closure of $D^+(S)$,
which is future development from $S$.
Thus, the data on $S$ predicts the entire future  for far away observers.
The other version, called the {\it strong cosmic censorship}, is a general
predictability requirement on any space-time, stating that all physically
reasonable space-times must be globally hyperbolic (see e.g. Penrose (1979)
in Ref. 1).
In effect, the weak cosmic censorship states that the region of space-time outside a black hole must be
globally hyperbolic.  A precise formulation
of this version of censorship will consist in specifying exact conditions
under which the space-time would be strongly asymptotically predictable.  In
its weak form the censorship conjecture does not allow
causal influences from singularity to asymptotic regions in space-time,
that is, singularity cannot be globally
naked.  However, it could be locally naked in the sense that  an observer
within the event horizon and in the interior of the black hole could possibly
receive particles or photons from the singularity.
Clearly, one needs to sharpen such a formulation.  For example, the metric
on $S$ should approach that of Euclidian three-space at infinity and
matter fields should satisfy suitable fall off conditions at spatial
infinity; also one might want the null generators of $\cal I^+$
to be complete, and one has to specify what exactly is meant by `physically
reasonable' matter fields.  In fact, as far as the cosmic censorship
hypothesis is concerned, it is a major problem in itself to find a
satisfactory and mathematically rigorous  formulation of what is physically
desired to be achieved.  Developing a suitable formulation would probably be
a major advance towards the solution of the main problem. It should be
noted that presently no general proof
is available for any suitably formulated version of the weak censorship. The
main difficulty appears to be that the event horizon is a feature depending
on the whole future behavior of the solution over an infinite time period,
whereas the present theory of quasi-linear hyperbolic equations guarantee
the existence and regularity of the solutions over a finite time internal
only.  In this connection, the results of Christodoulou [5] on
spherically symmetric collapse of a massless scalar field are relevant,
where it is shown using global existence theorems on partial differential
equations that global singularity free solutions can exist for weak enough
initial data.  In any
case, even if it is true, the proof for a suitable version of the weak
censorship conjecture would seem to require much more knowledge
on general global properties of Einstein's
equations and solutions than is known presently.

It is now possible to summarize the overall situation as follows.  Clearly, the
cosmic censorship hypothesis is a crucial assumption underlying all of
the black hole physics and gravitational collapse theory,
and several important related areas.
Whereas no proof for this conjecture is available, the first major task to be
accomplished here is in fact to formulate rigorously a satisfactory version
of the hypothesis.  The proof of cosmic censorship would confirm the already
widely accepted and applied theory of black holes, while its overturn would
throw the black hole dynamics into serious doubt.
Thus, cosmic censorship turns out to be one of the most important
open problem of considerable  significance for general
relativity and gravitation theory today.  Even if it is true, a proof for
this conjecture does not seem possible unless some major theoretical advances
by way of mathematical techniques and understanding the global structure of
Einstein equations are made.  In fact, the direction of theoretical advances
needed is not quite clear.  

This situation leads us to conclude that the first
and foremost task at the moment is to carry out a detailed and careful
examination of various gravitational collapse scenarios
to examine them for their end states.
It is such an investigation, of more general collapse
situations, which could indicate what theoretical advances to expect for
a proof and what features to avoid  while formulating the cosmic
censorship.
\bigskip

\noindent {\bf 3. Final Fate of Gravitational Collapse}
\smallskip

It would seem from the previous considerations that
we still do not have sufficient data and information
available on the various possibilities for gravitationally
collapsing configurations so as to decide one way or other on the
issue of censorship hypothesis. What appears really necessary is a detailed
investigation of different collapse scenarios, and to examine the
possibilities arising, in order to have
insights into the issue of the final fate of gravitational collapse.
It is with such a purpose that we would like to discuss now
several collapse situations involving different forms of matter to
understand the final fate of collapse.

Since we are interested in collapse scenarios,
we require that the space-time contains a regular initial spacelike
hypersurface on which the matter fields, as represented by the stress-energy
tensor $T_{ij}$, have a compact support and all physical quantities are
well-behaved on this surface. Also, we require the matter to
satisfy a suitable energy condition and that the Einstein equations are
satisfied. We say that the space-time contains a naked singularity
if there is a future directed non-spacelike curve which reaches a far
away observer or infinity in future, and in the past it terminates
at the singularity.

We will be mainly interested
in the nature of singularities arising as the final end product
of collapse, rather than for example, the {\it shell-crossing} naked
singularities which have been shown to occur in spherical collapse of
dust [6]
where shells of matter implode in such a way that
fast moving outer shells overtake the inner shells, producing a globally naked
singularity outside the horizon. These are singularities where
shells of matter pile up to give two-dimensional caustics and
the density and some curvature components blow up. The general point of
view is, however, such singularities need not be treated as serious
counter-examples to censorship hypothesis because these are merely consequent
to intersection of matter flow lines.  This gives a distributional
singularity which is gravitationally weak in the sense that curvatures and
tidal forces remain finite near the same.

On the other hand, there are  {\it shell-focusing} naked singularities
occurring at the center of spherically symmetric collapsing configurations
of dust or perfect fluid or radiation shells, as we shall consider here.
These are more difficult to ignore.  One can rule them out only by saying
that the dust or perfect fluids are not really `fundamental'
forms of matter.  However, if censorship is to be established as a
rigorous theorem, such objections have to be made precise in terms of a clear
and simple restriction on the stress-energy tensor, because these are forms
of matter which otherwise satisfy reasonability conditions such as the
dominant energy condition (provided there are no large negative pressures)
or a well posed initial value formulation for the coupled Einstein-matter
field equations. Further, these forms of matter
are widely used in discussing various astrophysical processes.

When should one regard a naked singularity forming in gravitational collapse
as a serious situation which must guide the formulation and proof of the
censorship hypothesis, or which must be regarded as an important
counter-example? The following could be imposed as a minimum set of conditions
for this purpose. Firstly, the naked singularity has to be visible at least
for a finite period
of time to any far away observer.  If only a single null geodesic escaped, it
would provide only an instantaneous exposure by means of a
single wave front.  In order to yield any observable consequences, a necessary
condition is that a family of future directed non-spacelike geodesics
should terminate at the naked singularity in past.  Next, this singularity
must not be gravitationally weak, creating a
mere space-time pathology, but must be a strong curvature singularity in a
suitable sense where densities and curvatures diverge sufficiently fast.
This would ensure that the space-time does not admit any continuous
extension through the singularity in any meaningful manner, and hence such
a singularity cannot be avoided. The physical effects due to strong fields
should then be important near such a strong curvature singularity.
Finally, the form of matter should be reasonable in that it must satisfy
a suitable energy condition ensuring the positivity of energy, and the
collapse ensues from an initial spacelike surface with a well-defined
non-singular initial data.

We first consider the phenomena of
gravitational collapse of a spherical shell of radiation in this context
and examine the nature and structure of resulting singularity with
special reference to censorship, and the occurrence of black holes and
naked singularities. The main motivation to discuss this situation is
this provides a clear picture in an explicit manner of what is possible in
gravitational collapse.

\smallskip
\noindent {\bf 3.1 The Vaidya-Papapetrou Radiation Collapse Models}
\smallskip

These are inflowing radiation space-times which represent the situation
of a thick shell of directed radiation collapsing at the center of symmetry
in an otherwise empty universe which is asymptotically flat far away.
The imploding radiation is described by the Vaidya space-time, given in
$(v,r,\theta,\phi)$ coordinates as
\be
ds^2=-\left(1 - {2m(v)\over r}\right)dv^2 + 2dvdr + r^2d\Omega^2
\ee
The radiation collapses at the origin of coordinates, $v = 0, r = 0$.
Throughout the discussion here the null coordinate $v$ denotes
the advanced time and $m(v)$ is an arbitrary but non-negative
increasing function of $v$.  The stress-energy tensor for the radial flux
of radiation is
\be
T_{ij} = \rho k_ik_j= {1 \over 4\pi r^2}{dm\over dv} k_ik_j
\ee
with $k_i = - {\delta^v}_i,\quad  k_ik^i = 0$,
which represents the radially inflowing radiation along the world lines
$v = \hbox{const.}$  Note that ${dm/dv} \ge 0$
implies that the weak energy condition is satisfied.  The situation
is that of a radially injected radiation flow into an
initially flat and empty region, which is focused into a central singularity
of growing mass by a distant source. The source is turned off at a finite
time $T$ when the field settles to the Schwarzschild space-time.  The Minkowski
space-time for $v < 0, m(v) = 0$ here is joined to a Schwarzschild space-time
for $v > T$ with mass $m_0 = m(T)$ by way of the Vaidya metric above.

The question is, will there be families of future
directed non-spacelike geodesics which might possibly terminate at the
singularity $v = 0, r = 0$ in the past, thus producing a naked
singularity of the space-time; and if so, under what conditions
this phenomena is ruled out to produce a black hole (see e.g. [2] and
references therein).
An interesting case, where many details can be explicitly worked out
to yield considerable insight into the final fate of these collapse models
is when the mass function $m(v)$ is chosen to be a linear function,
$2m(v) = \lambda v$, with $\lambda > 0$. This is the Vaidya$-$Papapetrou
space-time. In this case, $ 2m(v) = 0$  for $ v < 0$, $2m(v) = \lambda v$
for $0 < v < T$ and $ m(v) = m_0$ for $v > T$. Then, the
mass for the final Schwarzschild black hole is $m_{0}$ and the
causal structure of the space-time depends on the values chosen for the
constants $m_{0}, T,$ and $\lambda$. In this case the space-time admits
a homothetic Killing vector field and it is possible to work out all the
families of non-spacelike geodesics; and also
to determine when such future directed families will terminate at the
singularity in past, thus creating a naked singularity in the
space-time [7]. It turns out that it is the
time rate of collapse, as characterized by the value of the parameter
$\lambda$, which determines the formation of either a naked singularity
or a black hole as the end product of collapse in this case. For the
range $0<\lambda\le 1/8$, the singularity turns out to be naked with
families of infinitely many geodesics escaping away from the singularity.
For the range $\lambda>1/8$ the event horizon fully covers the singularity,
giving rise to a black hole.

The nature of this singularity, when it is naked, can be further
explored by examining the curvature growth along the families of geodesics
in the limit of approach to the singularity in the past.
One considers the behavior of scalars such as
$\psi = R_{ab} K^a K^b$, where $R^{ab}$ is the Ricci curvature
and $K^a$ is the tangent vector to the nonspacelike geodesics.
If $k$ is the affine parameter along the outgoing trajectories with
$k=0$ at the singularity, it is seen that $\psi$ diverges as
$1/k^2$ in the limit of approach to the singularity along $all$ the
families of nonspacelike geodesics coming out of the singularity.
This makes it the most powerful curvature singularity as characterized
in the literature (see e.g. Clarke in Ref. 1), which is as strong as the
big bang singularity of the Friedmann models.
Another important invariant that characterizes the strength of the
singularity is the Kretschmann scalar
near the naked singularity, which is given by
\be
\alpha = R^{ijkl}R_{ijkl} = {48 M(v)^2\over r^6}={12\lambda^2 X^2
\over r^6}
\ee
Examining the behavior of $\alpha$ along the families of the
non-spacelike geodesics joining the singularity, it is verified that
this scalar always diverges, thus establishing
a scalar polynomial singularity as expected. However, an interesting
directional behavior is revealed by the singularity as far as the scalar
$\alpha$ is concerned, which was not the case for the scalar $\psi$
above. Unlike $\psi$ whose behavior was independent of direction
of approach to singularity, the Kretschmann scalar not only shows directional
dependence but also a dependence on the parameter $\lambda$
which characterizes the rate at which the null dust is imploding.
Such a situation has been referred to as a `directional singularity', where
the singularity strength varies with direction. Ellis and King [8]
discussed such a directional property within the cosmological scenario
of Friedman models where the strength depends on the direction along which
the geodesic enters the singularity. Thus we see that similar
property is exhibited by the naked singularity resulting from
gravitational collapse.

To summarize, in the Vaidya$-$Papapetrou radiation collapse,
not just isolated radial null geodesics
but entire families of future directed non-spacelike geodesics
escape from the naked singularity in the past, which forms at the origin of
coordinates. The structure of these families and
the curvature growth along such trajectories illustrates that this is a
strong curvature visible singularity in a very powerful sense and
curvatures diverge very rapidly along all the families of
non-spacelike geodesics meeting the singularity in the past.
Thus, this is not a removable singularity. Further,
these results also
generalize to the case when the mass function is not linear and hence
space-time is not self-similar. Thus, the occurrence of naked singularity
cannot be attributed to the assumption of linearity of mass function.

\smallskip
\noindent {\bf 3.2 Inhomogeneous Dust Collapse}
\smallskip

How are the conclusions given in Section 2 for the
homogeneous dust collapse modified when the inhomogeneities of matter
distribution are taken into account?
It is important to include effects of inhomogeneities because
typically a realistic collapse would start from a very inhomogeneous
initial data with a centrally peaked density profile.
This problem can be investigated using the Tolman$-$Bondi-Lemaitre models [9],
which describe gravitational collapse of an inhomogeneous spherically
symmetric dust cloud, that is, a perfect
fluid with equation of state $p = 0$.  This is
an infinite dimensional family of asymptotically flat solutions of Einstein's
equations, which is matched to the Schwarzschild space-time outside
the boundary of the collapsing star. The Oppenheimer and Snyder [3]
homogeneous dust ball collapse is a special case of this class
of solutions.

This question of inhomogeneous dust collapse has attracted attention of many
researchers [10] and it is again seen that the introduction of inhomogeneities
leads to a rather different picture of gravitational collapse.  Shell-crossing
naked singularities occur in Tolman-Bondi-Lemaitre models when shells of dust
cross one another at a finite radius, which could be locally and even
globally naked. We shall not consider these for the reasons discussed earlier.
More important are the shell-focusing singularities occurring on the
central world line which we discuss below.

To discuss the structure of singularity forming in a
general class of Tolman-Bondi-Lemaitre models, the metric for spherically
symmetric collapse of inhomogeneous dust, in comoving
coordinates $(t, r, \theta,\phi)$, is given by,
\be
ds^2= -dt^2+{R'^2\over1+f}dr^2+R^2(d\theta^2+sin^2\theta\, d\phi^2)
\ee
\be
T^{ij}=\epsilon \delta^i_t \delta^j_t,\quad \epsilon=\epsilon(t,r)={F'
\over R^2R'}
\ee
where $T^{ij}$ is the stress-energy tensor, $\epsilon$ is
the energy density, and $R$ is a function of both $t$ and $r$ given by
\be
\dot R^2={F\over R}+f
\ee
Here the dot and prime denote partial derivatives with respect
to the parameters $t$ and $r$ respectively. As we are considering collapse,
we require $\dot R(t,r)<0.$ The quantities $F$ and $f$ are arbitrary
functions of $r$ and $4\pi R^2(t,r)$ is the proper area of the  mass shells.
The area of such a shell at $r=\hbox{const.}$ goes to zero when $R(t,r)=0$.
For gravitational collapse situation,
we take $\epsilon$ to have compact support
on an initial spacelike hypersurface and the space-time can be matched
at some $r=\hbox{const.}=r_c$ to the exterior Schwarzschild field
with total Schwarzschild mass $m(r_c)=M$ enclosed within the dust ball
of coordinate radius of $r=r_c$. The apparent horizon in the interior
dust ball lies at $R=F(r)$.

With the integration of equation for $\dot R$ above we have
in all three arbitrary functions of $r$, namely $f(r)$, $F(r)$, and $t_0(r)$
where the last indicates the time along the singularity curve.
One could, however, use the coordinate freedom left in the choice of
scaling of $r$ to reduce the number of arbitrary functions to two.
Thus, rescaling $R$ such that $R(0,r)=r$ leaves us with only two free
functions $f$ and $F$.
The time $t=t_0(r)$ corresponds to $R=0$ where the area of the shell of
matter at a constant value of the coordinate $r$ vanishes. It follows that
the singularity curve $t=t_0(r)$ corresponds to the time when the matter shells
meet the physical singularity. Thus, the range of coordinates is given by
$ 0\le r<\infty$, $ -\infty<t<t_0(r)$. It follows that unlike the collapsing
Friedmann case, or the homogeneous dust case, where the physical singularity
occurs at a constant epoch of time (say, at $t=0$), the singular epoch now
is a function of $r$ as a result of inhomogeneity in the matter distribution.
One could recover the Friedmann case from the above if we set
$t_0(r)=t'_0(r)=0.$ The function $f(r)$ classifies the space-time as bound,
marginally bound, or unbound depending on the range of its values
which are $f(r)<0, f(r)=0, f(r)>0$, respectively.
The function $F(r)$ is interpreted as the weighted mass
within the dust ball of coordinate radius $r$.
For physical reasonableness the weak energy condition is assumed,
that is, $T_{ij}V^iV^j\ge 0$ for all non-spacelike vectors $V^i$. This
implies that the energy density $\epsilon$ is everywhere positive,
($\epsilon \ge 0$) including the region near  $r=0$.
>From the scaling above, the energy density $\epsilon$ on the hypersurface
$t=0$ is written as $\epsilon={F'/r^2}.$ Then the weak energy condition
implies that $F'\ge0$ throughout the space-time.

Using the above framework, the nature of the singularity $R=0$ can
be examined. In particular, the problem of nakedness or otherwise of
the singularity can be reduced to the existence of real, positive roots
of an algebraic equation, constructed out of the free functions $F$ and
$f$ and their derivatives [11], which constitute the initial data of
this problem. We call the singularity to be a {\it central singularity}
if it occurs at $r=0$. Partial derivatives $R'$ and $\dot {R'}$
can be written as,
\be
\left({\partial R(t,r)\over \partial r}\right)_{t=\hbox{const.}}
=R'=(\eta-\beta)P-\left[ {1+\beta-\eta\over \sqrt{\lambda+f}}
+(\eta-\textstyle{3\over 2}\beta){t\over r}\right]\dot R
\ee
\be
\left({\partial R'(t,r)\over \partial t}\right)_{r=\hbox{const.}}=
{\beta\over 2r}\dot R +{\lambda\over 2rP^2}
\left[ {1+\beta-\eta\over \sqrt{\lambda+f}}
+(\eta-\textstyle{3\over 2}\beta){t\over r}\right]
\ee
where we have used the notation,
\be
R(t,r)=rP(t,r),\quad \eta=\eta(r)=r{F'\over F}, \quad \beta=\beta(r)=
r{f'\over f},\quad F(r)=r\lambda(r)
\ee
To focus the discussion, we restrict to functions
$f(r)$ and $\lambda(r)$ which are analytic at $r=0$ such that
$\lambda(0)\ne 0$.

The tangents $K^r=dr/dk$ and $K^t=dt/dk$ to the
outgoing radial null geodesics, with $k$ as the affine parameter, satisfy
\be
{dK^t\over dk}+{\dot R'\over\sqrt{1+f}}=0,\quad K^rK^t=0,\quad
{dt\over dr}={K^t\over K^r}={R'\over \sqrt {1+f}}
\ee
Our purpose is to find whether these geodesics terminate in the past at the
central singularity $r=0,t=t_0(0)$. The exact nature of this
singularity $t=0,r=0$ could be analyzed by the limiting
value of $X\equiv t/r$ at $t=0,r=0$. If the geodesics meet the singularity
with a definite value of the tangent then using l'Hospital rule we get
\be
X_0=\lim_{t\rightarrow 0,r\rightarrow 0}
{t\over r}=\lim_{t\rightarrow 0,r\rightarrow 0}{dt\over dr}=
\lim_{t=0,r=0}{R'\over \sqrt{1+f}}
\ee
where the notation is, $\lambda_0=\lambda(0),\beta_0=\beta(0),
f_0=f(0)$ and $Q=Q(X)=P(X,0)$.
Using the expression for $R'$ earlier, the above can be written as
$ V(X_0)=0,$ where
\be
V(X)\equiv (1-\beta_0)Q+ \left({\beta_0\over \sqrt{\lambda_0+f_0}}+
(1-{3\over2}\beta_0)X\right)\sqrt{{\lambda_0\over Q}+f_0} -
X\sqrt {1+f_0}
\ee
Hence if the equation $V(X)=0$ has a real positive root, the singularity
could be naked. In order to be the end point of null geodesics at least
one real positive value of $X_0$ should satisfy  the above.
Clearly, if no real positive root of the above is found, the singularity
$t=0,r=0$ is not naked. It should be noted that
many real positive roots of the above equation may exist which give the
possible values of tangents to the singular null geodesics
terminating at the singularity. However, such integral curves may or may
not realize a particular value $X_0$ at the singularity.
Suppose now $X=X_0$ is a simple root to $V(X)=0$.
To determine whether $X_0$ is realized as a tangent along any outgoing
singular geodesics to give a naked singularity, one can integrate the
equation of the radial null geodesics in the form $r=r(X)$
and it is seen that there is always atleast one null geodesic terminating
at the singularity $t=0,r=0$, with $X=X_0$. In addition there would be
infinitely many integral curves as well, depending on the values of the
parameters involved, that terminate at the singularity.
It is thus seen [11] that the existence of a positive real root of
the equation $V(X)=0$ is a
necessary and sufficient condition for the singularity to be
naked. Finally, to determine the curvature strength of the naked singularity
at $t=0$, $r=0$, one may analyze the quantity
$ k^2 R_{ab}K^aK^b$ near the singularity. Standard analysis shows that
the strong curvature condition is satisfied, in that the above
quantity remains finite in the limit of approach to the singularity.

The assumption of vanishing pressures here,
which could be important in the final stages of the collapse, may be
considered as the limitation of dust models.
On the other hand, it is also argued sometimes that in the final stages
of collapse, the dust equation of state could be relevant
and at higher and higher densities the matter may behave
more and more like dust. Further, if there are no large negative pressures
(as implied by the validity of the energy conditions), then the pressure
also might contribute gravitationally in a positive manner
to the effect of dust and may not alter the conclusions.
The considerations given to the case with non-zero
pressure are described below.

\smallskip
\noindent {\bf 3.3 Collapse with Non-zero Pressures}
\smallskip

It is clearly important to consider collapse situations which consider
matter with non-zero pressures and with reasonable equations of state.
It is possible that pressures may play an important role for the later
stages of collapse and one must investigate the possibility if pressure
gradients could prevent the occurrence of naked singularity.

We now discuss these issues, namely, the existence,
the termination of non-spacelike geodesic families, and the
strength of such a singularity for collapse with non-zero pressure.
Presently we consider only self-similar collapse models
and generalization of this will be discussed next. A numerical treatment
of self-similar perfect fluid was given by Ori and Piran [12] and the
analytic consideration was given by Joshi and Dwivedi [13].

A self-similar space-time is characterized by the existence of a homothetic
Killing vector field.  A spherically symmetric
space-time is self-similar if it admits a radial area coordinate $r$ and an
orthogonal time coordinate $t$ such that for the metric components
we have $g_{tt} (ct, cr) = g_{tt} (t, r), g_{rr} (ct, cr) = g_{rr} (t, r),$
for all $ c > 0$.  Thus, along the integral curves of the Killing vector
field all points are similar.  For the self-similar case, the
Einstein equations reduce to ordinary differential equations.

The spherically symmetric metric in comoving coordinates is given by,
\be
ds^2=-e^{2\nu(t,r)}dt^2 +e^{2\psi (t,r)}dr^2 +r^2S^2(t,r)
(d\theta^2+\sin^2\theta\, d\phi^2)
\ee
Self-similarity
implies that all variables of physical interest are expressed
in terms of the similarity parameter $X=t/r$ and so $\nu,\psi$
and $S$ are functions of $X$ only. The pressure and energy
density in comoving coordinates are ($u^a=e^{-\nu}\delta^a_t$)
\be
P={p(X)\over 8\pi r^2},\qquad \rho = {\eta(X)\over  8\pi r^2}
\ee
The field equations in this case for a
perfect fluid have been discussed in [14], and after
suitable integrations [13] these can be written as,
\be
e^{2\psi}= \alpha \eta^{{-2\over a+1}}S^{-4},\quad e^{2\nu}= \gamma
(\eta  X^2)^{{-2a\over a+1}}
\ee
\be
\dot V(X)=Xe^{2\nu}[(\eta +p)e^{2\psi}-2]=
Xe^{2\nu}[H-2]
\ee
\be
\left({\dot S \over S}\right)^2V+\left({\dot S  \over  S}\right)
(\dot V+ 2Xe^{2\nu})
+e^{2\nu+2\psi}\left(-\eta -e^{-2\psi}+{1\over S^2}\right)=0
\ee
where the dot denotes differentiation with respect to the similarity
parameter $X$. The quantities $V$ and $H$ here are defined by
\be
V(X) \equiv e^{2 \psi} - X^{2} e^{2 \nu},\quad H \equiv (\eta + p) e^{2 \psi}
\ee
Here the assumption is that the collapsing fluid is
obeying an adiabatic equation of state $p (X) = a \eta (X),$ with
$0 \le a \le 1$.  The special case $a = 0$
describes dust and $a = 1/3$ gives the equation of state for radiation.

The point $t = 0$, $r = 0$, is a singularity where the density
necessarily diverges.  Such a divergence is also
observed when we approach the singularity along any line of self-similarity
$X = X_{0}$.  This leads to the divergence of curvature scalars such as
${R^j}_{i} {R^i}_{j}$ and also of the Ricci scalar $R = \rho - 3p$ if
$a \not = 1/3$.  It is to be examined whether this singularity could be
possibly naked, and if so whether families of non-spacelike geodesics would
terminate at the same in the past.

For this purpose, the first requirement is, the families of
non-spacelike geodesics are to be integrated. This is possible in this
case using the existence of a homothetic Killing vector field [13].
The geodesic equations are then,
\be
{dt\over dr}={X\pm e^{2\psi} Q\over 1\pm Xe^{2\nu}Q}
\ee
For a specific discussion, let us choose the function $Q$ to be
positive throughout and $\pm$ signs represent outgoing or ingoing solutions.
The point $t=0,r=0$, is a singular point of the above differential
equation. The nature of the limiting value of
$X=t/r$ plays an important role in the analysis of non-spacelike
curves that terminate at the singularity in past and reveals the exact
structure of the singularity. From geodesic equations, using l'Hospital's rule
we get
\be
X_0=\lim_{t\to 0,r\to 0}{t\over r}=\lim_{t\to 0,r\to 0}{dt\over dr}=
{X_0\pm e^{2\psi(X_0)}Q(X_0)\over 1\pm X_0e^{2\nu(X_0)}Q(X_0)}
\ee
Thus we see that if trajectories are to go out of the singularity, then
there exists a real positive value $X_0$ such that
\be
V(X_0)\equiv e^{2\psi(X_0)}-X_0^2e^{2\nu(X_0)}=0
\ee
If $V(X_o)=0 $ has no real positive roots then geodesics clearly do not
terminate at the singularity with a definite tangent.
In case when
$V(X_0)=0$ has a positive real root, a single
radial null geodesic would escape from the singularity.
Existence of positive real roots of $V(X)=0$ is therefore
a necessary and sufficient condition for the singularity to be
naked and at least one single null geodesic in the ($t,r$) plane would escape
from the singularity.

In order to find whether a family of null or timelike geodesics would
terminate at the singularity in the present case,
one can consider the equation of geodesics $r=r(X)$ in $(r,X)$ plane,
restricting to positive sign solutions which are outgoing geodesics.
Suppose $V(X)=0$ has one simple real positive root $X=X_0$. Then using
the equation for $\dot V(X)$ given above, we can write near the singularity
$V(X)=(X-X_o)X_oe^{2\nu(X_o)}(H(X_o)-2)$,
and using the fact that $Q$ is positive, geodesics can be integrated
near the singularity to get
\be
r=D(X-X_o)^{2\over H_o-2}
\ee
where $H_o=H(X_o)$. When $H_o>2$ it is seen that an infinity of
integral curves will meet the singularity in the past with tangent $X=X_o$;
different curves being characterized by different values of the constant $D$.
It follows that this singularity is at least locally naked from which an
infinity of non-spacelike curves are ejected.
In the case $H_o<2$ but $H_o> 0$, the singularity
would be a node in the $(r,t)$ plane and infinity of curves will still
escape. However, if $H_o<0$, in $(r,t)$ plane the
integral curves move away from the singularity and never terminate
there. Thus, we see that infinitely many integral curves would terminate
at the singularity as long as
\be
\infty > H_o=H(X_o)=(\eta+p)e^{2\psi }> 0
\ee
The above will be satisfied provided the weak energy
condition holds and further
that the energy density as measured by any timelike observer is positive
in the collapsing region near the singularity. One then examines
the curvature strength of the singularity along the
trajectories coming out and it is again seen that the strong curvature
condition is satisfied.

The results could be summarized as follows. If in a self-similar
collapse a single null radial geodesic escapes the singularity,
then an entire family of non-spacelike geodesics would also escape provided
the positivity of energy density
is satisfied as above. It also follows that no families of non-spacelike
geodesics would escape the singularity, even though a single null
trajectory might, if the weak energy condition is violated.

\smallskip
\noindent {\bf 3.4 Gravitational Collapse with General Form of Matter}
\smallskip

Consideration of matter forms above such as directed radiation, dust,
perfect fluids etc imply some general pattern emerging about the final
outcome of gravitational collapse. Hence one could ask the question
whether the final fate of collapse would be independent of the form of
matter under consideration. An answer to this is important because it
has often been conjectured in the literature (see e.g. [1,2])
that once a suitable form of matter with an appropriate equation of state,
and satisfying energy condition, is considered then there may be no
naked singularities. After all, there is always a possibility that
during the final stages of collapse matter may not have any of the
forms considered above, because such relativistic fluids are
phenomenological and one must treat matter in terms of fundamental fields,
such as for example, a massless scalar field.

Some efforts in this direction are worth mentioning where the above
results on perfect fluid were generalized to matter
forms without any restriction on the form of $T_{ij}$, which was
supposed to satisfy the weak energy condition only [15]. A consideration
to a general form of matter was also given by Lake; and by Szekeres
and Iyer [16] who do not start by assuming an equation of state but
a class of metric coefficients is considered with a certain power law
behavior. The main argument of the results such as [15] is along the following
lines. In the discussion above it was pointed out that naked
singularities could form in the gravitational collapse from a regular initial
data, from which non-zero measure families of non-spacelike trajectories
come out. The criterion for the existence of such singularities
was characterized in terms of the existence of real positive roots of
an algebraic equation constructed out of the field variables. A similar
procedure was developed now for general form of matter. In comoving
coordinates, the general matter can be described by three functions,
namely the energy density and the radial and tangential pressures. The
existence of naked singularity is again characterized in terms of the
real positive roots of an algebraic equation, constructed from the
equations of non-spacelike geodesics which involve the three metric
functions. The field equations then relate these metric functions to the
matter variables and it is seen that for a subspace of this free initial data
in terms of matter variables, the above algebraic equation will have
real positive roots, producing a naked singularity in the space-time.

It thus follows that the occurrence of naked singularity is basically
related to the choice of initial data to the Einstein field equations,
and would therefore occur from  regular initial data within
the general context considered, subject to the matter satisfying
weak energy condition. The condition on initial data which leads to the
formation of black hole is also characterized in these works.
It then appears that the occurrence of naked singularity or a black hole
is more a problem of choice of the initial data for field equations
rather than that of the form of matter or the equation of state.
This has important implication for cosmic censorship in that in order to
preserve the same one has to avoid all such regular initial data causing
naked singularity, and hence a deeper understanding of the initial data
space is required in order to determine such initial data and the kind
of physical parameters they would specify. This would, in other words,
classify the range of physical parameters to be avoided for a particular
form of matter. More importantly, it would also pave the way for the
black hole physics to use only those ranges of allowed parameter values which
would produce black holes, thus putting black hole physics on a more
firm footing.

\smallskip
\noindent {\bf 3.5 Scalar Field Collapse}
\smallskip

Much attention has been devoted in past years to analyze the
collapse of a scalar field, both analytically [5, 18], as well as more
recently numerically [17]. We now discuss this case briefly. This is a
model problem of a single massless
scalar field which is minimally coupled to gravitational field and
it provides possibly one of the simplest scenarios to investigate the
nonlinearity effects of general relativity. On the analytic side, the
results by Christodoulou show that when the scalar field is sufficiently
weak, there exists a regular solution, or global evolution for an
arbitrary long time of the coupled Einstein and scalar field equations.
During the collapse, there is a convergence towards the origin, and
after a bounce the field disperses to infinity. For strong enough field,
the collapse is expected to result into a black hole. For self-similar
collapse, he has also obtained results to show that the collapse will
result into a naked singularity. However, he claimed that the initial
conditions resulting into a naked singularity are a set of measure zero
and hence the naked singularity formation may be an unstable phenomenon.

Such an approach helps study the cosmic censorship problem as the
evolution problem in the sense of examining the global Cauchy
development of
a self-gravitating system outside an event horizon. A `dynamical'
version of cosmic censorship will demand that given reasonable initial data
which is asymptotically flat, and assuming some reasonable energy
conditions, there exists a global Cauchy evolution of the system
outside the event horizon in the
sense that the solution exists for arbitrary large times for an asymptotic
observer. For a discussion of such an approach in the context of
self-gravitating scalar fields we refer to Malec in Ref. 18. He considers the
problem of global existence of solutions and also finds an explicit
example of an initial configuration that results into a naked singularity
at the center of symmetry.

The problem of scalar field collapse has been numerically studied by
Choptuik and others [17]. Choptuik considered a family of scalar field
solutions where a parameter $p$ characterized the strength of the
scalar field. His numerical calculations showed that for black hole
formation, there is a critical limit $p\to p^*$ and the mass of the
resulting black holes satisfy a power law $M_{bh}\propto (p-p^*)^\gamma$,
where the critical exponent $\gamma$ has value of about 0.37.
It was then conjectured that such a critical behavior may be a general
property of gravitational collapse, because similar behavior was found
by Abraham and Evans for imploding axisymmetric gravitational waves.
Also, Evans and Coleman considered the collapse of radiation with equation
of state $p=\rho/3$, assuming self-similarity for solutions.
It is still not clear if the critical parameter $\gamma$ will have the
same value for all forms of matter chosen and some further investigation
may be required to determine this issue. As the parameter $p$ moves from
the weak to strong range, very small mass black holes can form. This has
relevance to censorship, because in such a case one can probe and receive
messages from arbitrarily near to the singularity and this is naked
singularity like behavior. Attempts have also been made to construct
models analytically which may reproduce such a critical behavior [18],
assuming self-similarity. In particular, Brady has constructed solutions which
have dispersal, and also solutions with black holes or naked singularities.

\smallskip
\noindent {\bf 3.6 Non-spherical Collapse}
\smallskip

What will be the final fate of gravitational collapse which is not
spherically symmetric? The main phases of spherical collapse of a
massive star would be typically instability,
implosion of matter, and subsequent formation of an event horizon and
a space-time singularity of infinite density and curvature with infinite
gravitational tidal forces. This singularity may or may not be fully
covered by the horizon as we have discussed above.
Again, small perturbations over the spherically symmetric situation
would leave the situation unchanged [19] in the sense that an event horizon
will continue to form in the advanced stages of the collapse.

The next question is, do horizons still form when the fluctuations from the
spherical symmetry are high and the collapse is highly non-spherical?
It was shown by Thorne [20], for example, that when there is no spherical
symmetry, the collapse of infinite cylinders do give rise to naked
singularities in general relativity, which are not covered by horizons.
This situation motivated Thorne to propose the following {\it hoop conjecture}
for finite systems in an asymptotically flat space-time, which
characterizes the final fate of non-spherical collapse: The
horizons of gravity form when and only when a mass $M$ gets compacted in a
region whose circumference in {\it every} direction obeys
$ {\cal C}\le 2\pi(2GM/c^2)$. Thus, unlike the cosmic censorship conjecture,
the hoop conjecture does not rule out {\it all} the naked singularities
but only makes a definite assertion on the occurrence of the event
horizons in gravitational collapse. We also note that the hoop conjecture
is concerned with the formation of event horizons, and not with naked
singularities. Thus, even when event horizons form, say for example in
the spherically symmetric case, it does not rule out the existence of
naked singularities, i.e. it does not imply that such horizons must
always cover the singularities.

When the collapse is sufficiently aspherical, with one or two dimensions
being sufficiently larger than the others, the final state of collapse could
be a naked singularity, according to the hoop conjecture. Such a situation
is inspired by the Lin, Mestel and Shu [21] instability in Newtonian
gravity where a non-rotating homogeneous spheroid collapses, maintaining
its homogeneity and spheroidicity but its deformations grow. If the initial
condition is that of a slightly oblate spheroid, the collapse results into a
pancake singularity through which the evolution could proceed. However, for a
slightly prolate spheroidal configuration, the matter collapses to a thin
thread which ultimately results into a spindle
singularity. This is more serious in nature in that the gravitational
potential and the tidal forces blow up as opposed to only density
blowing up in a shell-crossing singularity. Even in the case of an
oblate collapse, the passing of matter through the pancake causes prolateness
and subsequently a spindle singularity again results without the
formation of any horizon.

It was indicated by the numerical
calculations of Shapiro and Teukolsky [22] that a similar situation
maintains in general relativity as well in conformity with the hoop conjecture.
They evolved collissionless gas spheroids in full general relativity which
collapse in all cases to singularities. When the spheroid is sufficiently
compact a black hole may form, but when the
semimajor axis of the spheroid is sufficiently large, a spindle singularity
results without an apparent horizon forming. One could treat these only
as numerical results, as opposed to a full analytic treatment,
which need not be in contradiction to a suitably formulated version of
cosmic censorship. However, this gives rise to the possibility of
occurrence of naked singularities in collapse of finite
systems in asymptotically flat space-times, which could be in violation
of weak cosmic censorship but in conformity with the hoop conjecture.

Apart from such numerical simulations, some analytic treatments of
aspherical collapse are also available. For example, the aspherical
Szekeres models for irrotational dust without any Killing vectors,
generalizing the spherical Tolman-Bondi-Lemaitre collapse, were recently
studied [23] to deduce the existence of strong curvature central naked
singularities.
While this indicates that naked singularities are not necessarily confined
to spherical symmetry, it must be noted that dynamical evolution
of a non-spherical collapse still remains a largely uncharted territory.

\bigskip

\noindent {\bf 4. Concluding Remarks and Open Issues}
\smallskip

We have discussed here several gravitational collapse scenarios
and the following pattern appears to be emerging. In the first place,
singularities not covered fully by the
event horizon, i.e. naked singularities, could occur in several collapsing
configurations from regular initial data, with reasonable
equations of state such as describing radiation, dust or a perfect fluid
with a non-zero pressure, or also for general forms of matter.  Secondly,
families of a non-zero measure set of photons or particles, escape from
such a naked singularity (which is a region of extreme densities in the
space-time) to reach far-away observers. Finally, the singularity is
physically significant in that densities and curvatures diverge powerfully
near such a naked singularity. It would appear that such results regarding
the final fate of gravitational collapse, generated from study of
different physically reasonable collapse scenarios, may provide useful
insights into black hole physics and may be of help for any possible
formulation of the cosmic censorship hypothesis.

One possible insight, for example, could be that the final
state of a collapsing star, in terms of either a black hole or a naked
singularity, may not really depend on the form or equation of state
of collapsing matter, but is actually determined by the physical initial
data in terms of the initial density profiles and pressures.
As a specific example, the role played by the initial density
and velocity distributions was examined [24] for the gravitational collapse of
spherically symmetric inhomogeneous dust cloud. The collapse can end in
either a black hole or a naked singularity depending on the values of
initial parameters.
In the marginally bound case, the collapse ends in a naked singularity
if the leading nonvanishing derivative at the center is either the first
one or the second one. If the first two derivatives are zero and the third
derivative non-zero, the singularity is naked and strong, or covered,
depending on a quantity determined by the third derivative and the central
density. If the first three derivatives are zero, the collapse ends in a
black hole. Analogous results are found when the cloud is not marginally
bound, and also for a cloud for which the collapse begins from rest.
There is a transition from the naked singularity phase to the black hole
phase as the initial density profile is made more and more homogeneous near
the center. As one progresses towards more homogeneity (and hence towards
a stronger gravitational field), there first occurs a weak naked singularity,
then a strong naked singularity, and finally a black hole.

The important question then is the genericity and stability of such
naked singularities arising from regular initial data. Will the initial
data subspace, which gives rise to naked singularity as end state
of collapse, have zero measure in a suitable sense? In that case, one
would be able to reformulate more suitably the censorship hypothesis,
based on a criterion that naked singularities
could form in collapse but may not be generic.

It has also been proposed (see e.g. [25]) that one may try to evolve
some kind of
a physical formulation for cosmic censorship, where the available studies
on various gravitational collapse scenarios such as above may provide a
useful guide. The idea here would be to study the various properties
of naked singularities collectively as they emerge from the studies so far.
One would then argue that objects with such properties are not physical
and hence would not matter in any manner even if they occurred in nature.
We list some of the possibilities in this direction below:

(i) One may, for example, ask about the `mass' of such a naked
singularity. If they turned out to be massless with a suitably
well-defined meaning of mass, then they may not be physically treated as
important. One should keep in view here the difficulties associated
with specifying notions such as mass and energy in general relativity.

(ii) Another possibility is to argue that the redshift from such a naked
singularity would be infinite and hence it would not be observable any way.
Here also, we have to define redshift carefully, as we are no longer
dealing with a regular event of the space-time, but with the redshift
of a singularity where densities and curvatures diverge.

(iii) It is possible that the use of a different system, such as the
Einstein-Vlasov equations, in order to trace the evolution of collapse
might help avoid naked singularities, or for that matter singularities
of all kinds. Some investigations in this direction have been reported [26]
and this alternative needs to be pursued further. Studying exact solutions
of the Einstein-Maxwell equations describing moving black holes, or further
investigation of available static exact solutions also might yield interesting
insights on the nature and properties of naked singularities and may
provide a test for censorship [27].

(iv) Still another possibility is to invoke
quantum effects and quantum gravity. While naked singularities may form
in classical general relativity, quantum gravity should presumably remove
them. So, why bother about them? The point here is that even though the
final singularity may be removed in this way, still there would be very
high density and curvature regions in the classical regime,
which would be causally
communicating with outside observers, as opposed to the black hole situation
where this is not the case. Some interesting efforts have been made
recently to show that quantum effects could remove the naked singularity [28];
this would then be a {\it quantum cosmic censorship}.

\bigskip
\bigskip
\noindent{\bf References}

\smallskip
\begin{description}

\item{[1]} R. Penrose, Riv. del. Nuovo Cim. 1, 252 (1969); Nature 236, 377
(1972); S. W. Hawking, Gen Relativ. Grav. 10, p.1047 (1979); R. Penrose,
in {\it General Relativity- an Einstein Centenary Survey} (eds. S. W. Hawking
and W. Israel), Cambridge University Press, Cambridge (1979); W. Israel,
Found. Phys. 14, p.1049 (1984); C. J. S. Clarke, {\it `Analysis of
space-time singularities'}, Cambridge University Press, Cambridge (1993).

\item{[2]} P. S. Joshi, {\it `Global aspects in gravitation and cosmology'},
Clarendon Press, OUP, Oxford (1993).

\item{[3]} J. Oppenheimer and H. Snyder, Phys. Rev. 56, p.455 (1939).

\item{[4]} S. W. Hawking and G. F. R. Ellis, {\it The large scale
structure of space-time}, Cambridge University Press, Cambridge (1973).

\item{[5]} D. Christodoulou, Commun. Math. Phys. 93, p.587 (1986);
Christodoulou, Commun. Math. Phys. 109, p.591; 109, p.613 (1987);
Commun. Pure and Applied Math. XLIV, p.339 (1991); Ann. Math. 140,
p.607 (1994).

\item{[6]} P. Yodzis, H.-J. Seifert and H. Muller zum Hagen,
Commun. Math. Phys. 34, p.135 (1973).

\item{[7]} I. H. Dwivedi and P. S. Joshi, Class. Quantum Grav.
6, 1599 (1989); 8, 1339 (1991).

\item{[8]} G. F. R. Ellis and A. R. King, Commun. Math. Phys. 38, p.119 (1974).

\item{[9]} R. C. Tolman, Proc. Natl. Acad. Sci. USA, 20, p.410 (1934);
 H. Bondi, Mon. Not. Astron. Soc., 107, p.343 (1947).

\item{[10]} D. M. Eardley and L. Smarr, Phys. Rev. D 19, p.2239 (1979);
D. Christodoulou, Commun. Math. Phys., 93, p.171 (1984); R. P. A. C. Newman,
Class. Quantum Grav., 3, p.527 (1986); B. Waugh and K. Lake, Phys. Rev. D 38,
p.1315 (1988); 
I. H. Dwivedi
and P. S. Joshi, Class. Quantum Grav., 9, L69 (1992); P. S. Joshi and I. H.
Dwivedi, Phys. Rev. D 47, p.5357 (1993);  
T.P. Singh and P. S. Joshi, Class. Quantum Grav.,
13, p.559 (1996); 
S. Jhingan, P. S. Joshi and T. P. Singh,
gr-qc/9604046 (1996).

\item{[11]} P. S. Joshi and I. H. Dwivedi, Ref. 10 above.

\item{[12]} A. Ori and T. Piran, Phys. Rev. D 42, p.1068 (1990).

\item{[13]} P. S. Joshi and I. H. Dwivedi, Commun. Math. Phys.
146, p.333 (1992); Lett. Math. Phys. 27, p.235 (1993).

\item{[14]} M. E. Cahill and A. H. Taub, Commun. Math. Phys. 21, p.1 (11971).

\item{[15]} I. H. Dwivedi and P. S. Joshi, Commun. Math. Physics,
166, p.117 (1994).

\item{[16]} K. Lake, Phys. Rev. Lett. 68, p.3129 (1992); P. Szekeres
and V. Iyer, Phys. Rev. D 47, p.4362 (1993).

\item{[17]} M. W. Choptuik, Phys. Rev. Lett. 70, p9 (1993); A. M. Abrahams
and C. R. Evans, Phys. Rev. Lett. 70, p.2980 (1993); C. R. Evans and
J. S. Coleman, Phys. Rev. Lett. 72, p.1782 (1994).

\item{[18]} M. D. Roberts, Gen. Relat. Grav. 21, p.907 (1989); J. Traschen,
Phys. Rev. D 50, p.7144 (1994); P. R. Brady, Class. Quant. Grav.
11, p.1255; Phys. Rev. D 51, p.4168 (1995); C. Gundlach, Phys. Rev. Lett.
75, p.3214 (1995); E. Malec, gr-qc/9506005, (1995).

\item{[19]} A. G. Doroshkevich, Ya. B. Zel'dovich and I. D. Novikov,
Sov. Phys. JETP, 22, p.122 (1966); V. de la Cruz, J. E. Chase and W. Israel,
Phys. Rev. Lett. 24, p.423 (1970); R. H. Price, Phys. Rev. D 5, p.2419 (1972).

\item{[20]} K. S. Thorne, in {\it `Magic without magic: John Archibald Wheeler'},
(ed. John R. Clauder), W. H. Freeman (1972).

\item{[21]} C. Lin, L. Mestel and F. H. Shu, Ap. J. 142, p.1431 (1965).

\item{[22]} S. L. Shapiro and S. A. Teukolsky, Phys. Rev. Lett. 66,
p.994 (1991); Phys. Rev. D 45, p.2006 (1992).

\item{[23]} P. S. Joshi and A. Krolak, gr-qc/9605033, (1996).

\item{[24]} P. S. Joshi and T. P. Singh in Ref. 14 above; for a further
discussion on this and related issues, see also T. P. Singh,
gr-qc/9606016 (1996).

\item{[25]} P. S. Joshi and I. H. Dwivedi in Ref. 14 above, Section V.

\item{[26]}A. D. Rendall, Class. Quant. Grav. 9, L99 (1992);
G. Rein, A. D. Rendall and J. Schaeffer, Commun. Math. Phys. 168, p.467 (1995).

\item{[27]} D. R. Brill, gr-qc/9501023 (1995); K. S. Virbhadra,
gr-qc/9606004 (1996).

\item{[28]} C. Vaz and L. Witten, Phys. Lett. B 325, p.27 (1994);
Class. quant. Grav. 12, p.2607 (1995); gr-qc/9511018 (1995);
gr-qc/9604064 (1996).

\end{description}

\end{document}